\documentstyle[aps,epsf,12pt]{revtex}
\tighten
\begin{document}
\title{Hierarchy in chaotic scattering in Hill's problem}
\author{Zolt\'an Kov\'acs\protect\thanks{Permanent address:
          Institute for Theoretical Physics, 
          E\"otv\"os University, Budapest}}
\address{Instituut voor Theoretische Fysica, Universiteit van Amsterdam\\
         Valckenierstraat 65, NL--1018 XE Amsterdam, The Netherlands}
\maketitle

\begin{abstract}
Hierarchic properties of chaotic scattering in a model of satellite
encounters, studied first by Petit and H\'enon, are examined by
decomposing the dwell time function and comparing scattering trajectories.
The analysis reveals an (approximate) ternary organization in the chaotic
set of bounded orbits and the presence of a stable island. 
The results can open the way for a calculation of global quantities
characterizing the scattering process by using tools of the
thermodynamic formalism.

\end{abstract}

\section{Introduction}
\label{sec-intro}

The idea that chaotic scattering \cite{chat-intro} may play an important
role in various problems in celestial mechanics became widely accepted
after the pioneering work of Petit and H\'enon \cite{Petit-Henon}.
In their study of satellite encounters (also known as Hill's problem),
they provided evidence for the presence of chaotic scattering by analyzing
the dependence of a suitable parameter characterizing the outgoing
asymptotic motion on the same parameter in the incoming asymptotics.
They pointed out that the irregular behaviour of this scattering function,
showing singularities on a fractal set, reflects the existence of
scattering trajectories that are asymptotic to bounded orbits of the system.

The bounded orbits form a fractal set---called the chaotic set or
nonattracting invariant set \cite{BGO}---in phase space with stable
manifolds extending smoothly into the region of asymptotic motion.
Whenever the initial condition of a scattering trajectory is placed on
a stable manifold, it gets trapped in the corresponding bounded orbit, 
so the structure of the singularities in the scattering function 
reflects the fractal structure of the chaotic set. 
By presenting trajectory plots of various types, Petit and H\'enon
underlined this structure but attempted no detailed characterization of it. 
In this paper, we provide a hierarchic analysis of chaotic scattering in
Hill's problem by studying a slightly different scattering function,
giving the dependence of the dwell time on the initial conditions,
combined with a careful examination of scattering orbits.
This method has recently been applied to a simple model of chemical
reactions \cite{KZLW} making properties of the hierarchic organization
clear in that problem.

At first, we give a brief introduction in Sec.~\ref{sec-hill} to Hill's
problem based on Ref.~\cite{Petit-Henon}, then we present the 
dwell time function and the basic ideas of its hierarchic analysis 
in Sec.~\ref{sec-dwtime}.
In Sec.~\ref{sec-orbits}, a selection of scattering orbits is used to
point out the presence of a ternary organization in the scattering data
and, correspondingly, in the structure of the chaotic set.
Finally, in Sec.~\ref{sec-discuss}, we discuss the consequences of our
findings including the possible use of the hierarchic information in the
calculation of average quatities in Hill's problem or other models 
exhibiting chaotic scattering.

\section{Satellite encounters}
\label{sec-hill}

The behaviour of two small bodies (satellites) moving along coplanar
circular orbits around a large central object (the planet) in the same
(counterclockwise) direction can approximately be described, 
in each other's vicinity, 
by Hill's equations \cite{Petit-Henon}:
\begin{eqnarray}
     \ddot{\xi}  & = &  2 \dot{\eta} + 3 \xi - \frac{\xi}{\rho} \\
     \ddot{\eta} & = & -2 \dot{\xi} - \frac{\eta}{\rho}
\end{eqnarray}
with $\rho = \sqrt{\xi^2 + \eta^2}$.
The coordinates $\xi$ and $\eta$ describe the positions of the satellites 
relative to each other, after an appropriate change of scales,
in a comoving reference frame rotating around the planet.
(In particular, if one of the satellites is much smaller than the other, 
then these coordinates simply give the position of the small
satellite while the big one rests in the origin.)
These equations contain no free parameter.
They can be derived as the Hamiltonian equations of motion from a suitable
function $H$, and, as a consequence, they give rise to a conserved quantity 
\begin{equation}
     \Gamma = 3 \xi^2 + \frac{2}{\rho} - \dot{\xi}^2 - \dot{\eta}^2
\end{equation}
playing the role of energy in the problem.

Hill's equations allow general coplanar motions of the satellites, 
not just circular orbits. 
Neglecting the interaction between the satellites, the ``free'' orbits
can be parametrized by the difference $h$ of the two semimajor axes and a
parameter $k$ called the reduced eccentricity; for circular obits 
$k=0$ and $h$ is just the difference of the two orbit radii. 
Therefore, the family of circular initial orbits can be parametrized by a
single parameter $h$ (apart from a trivial constant that can be transformed 
out by a suitable choice of the origin of time), and the constant of motion 
$\Gamma$ can be expressed through $h$: $\Gamma = \frac{3}{4} h^2$.
The symmetries of the equations also allow us to assume $h > 0$.
We will concentrate on this situation in the following and leave the case
of more general initial conditions to the discussion. 

If $h \neq 0$, the satellites slowly approach and get close to each other
(i.e., they have an encounter) due to the difference in their periods.
During the encounter, their interaction must be taken into account; 
as a result, they may perform complicated motion in the proximity of each
other for a while, but then separate again and move away from each other
asymptotically along generic elliptic orbits characterized by parameters
$h'$ and $k'$.
Bearing in mind the Hamiltonian character of the problem, this process can
be thought of as a scattering event from an incoming asymptotics
with a given value of $h$ and $k=0$ into an outgoing asymptotics 
characterized by $h'$ and $k'$.
The complicated motion during the encounter, which gives the chaotic
features to the scattering, is a manifestation of transient chaos \cite{Tel}, 
i.e., chaotic motion on a finite time scale.
In Ref.~\cite{Petit-Henon}, the function $h'(h)$ was used to point out the
presence of chaotic scattering; in the next section, we use a nowadays
standard approach based on the dwell time (or time delay) function.

\section{The dwell time function}
\label{sec-dwtime}

\begin{figure}
\centering\leavevmode \epsfysize=10cm\epsfbox{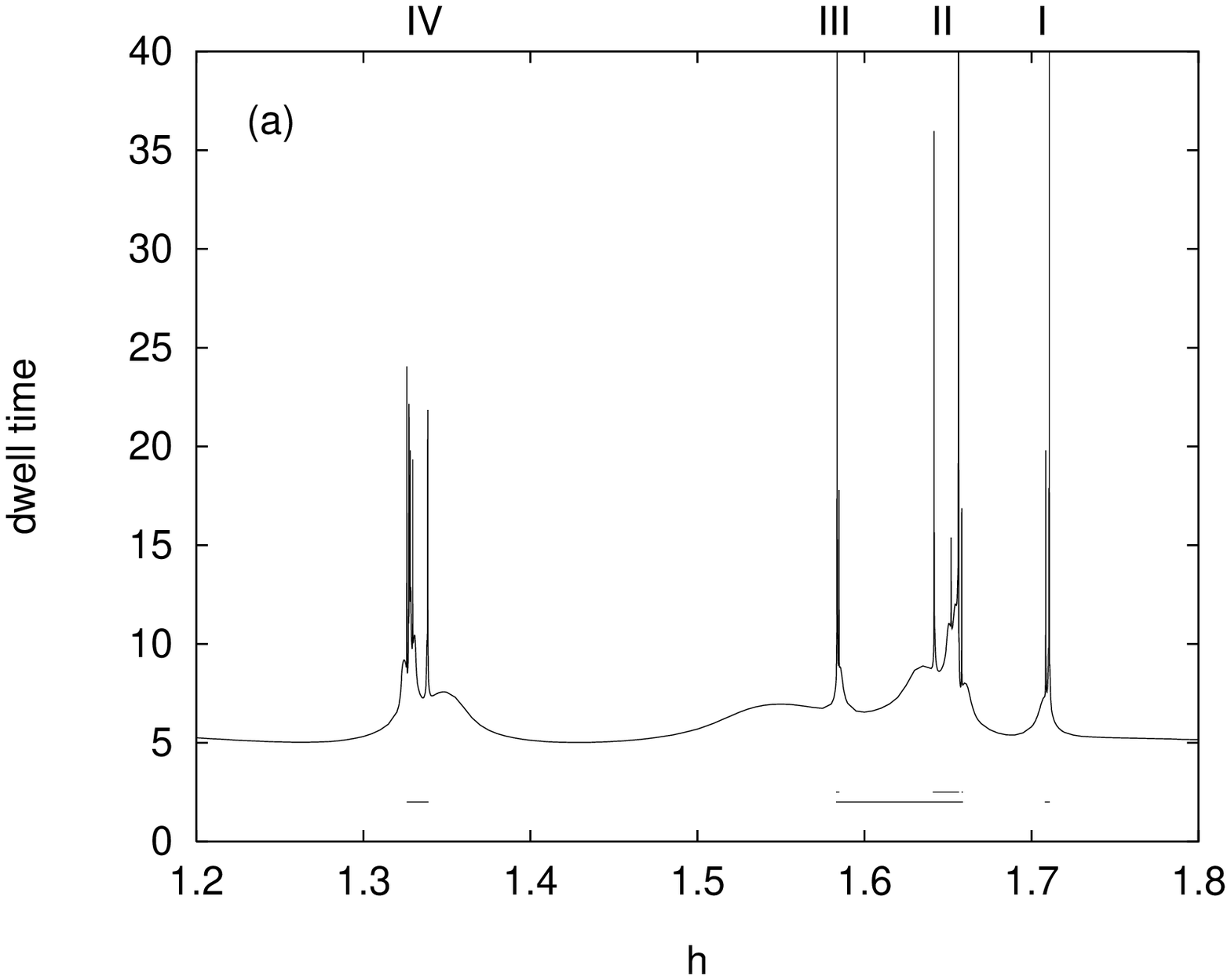} \\
\centering\leavevmode \epsfysize=10cm\epsfbox{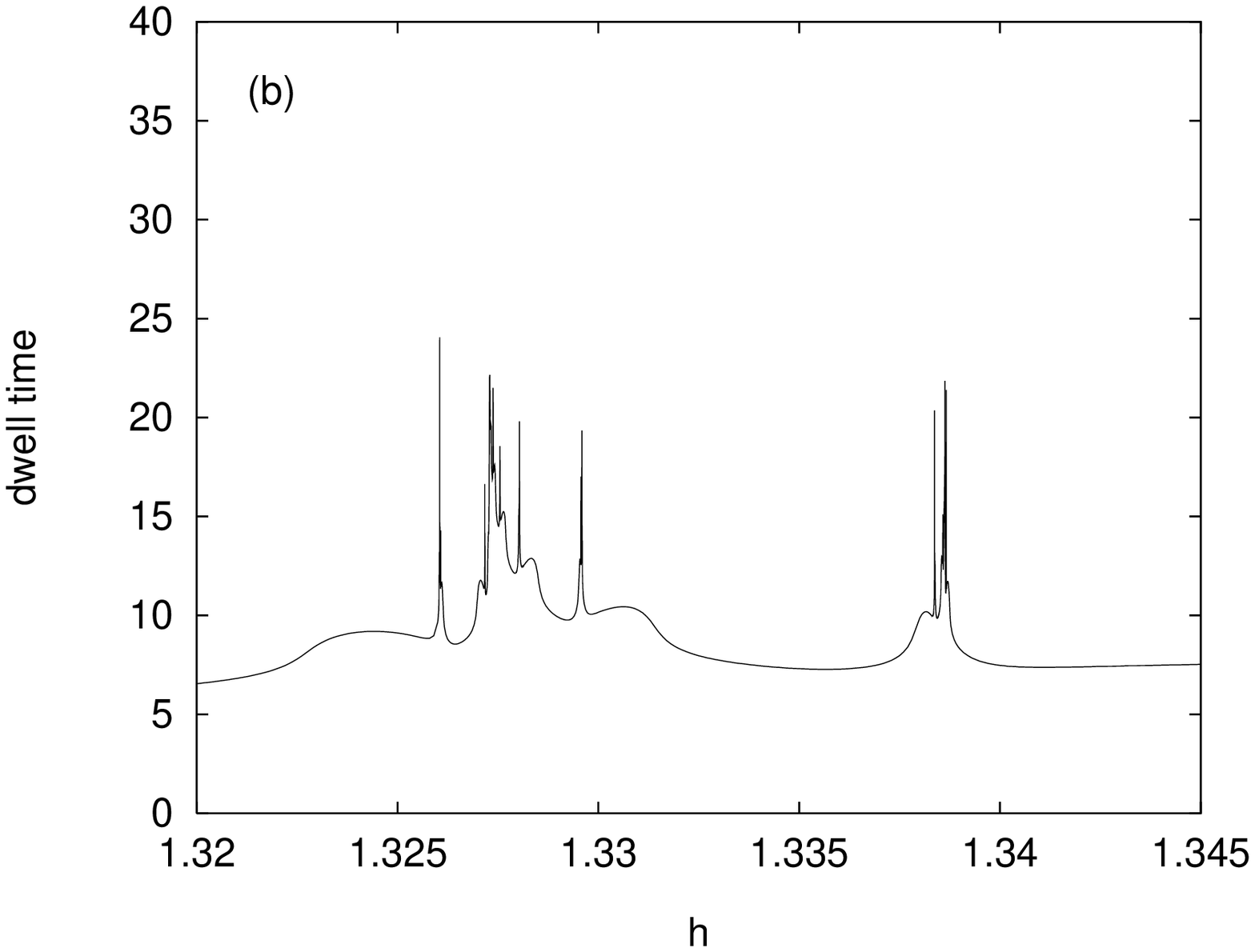} \\[.2cm]
\caption{%
(a) The dwell time function $T(h)$ for circular initial orbits with 
    $\eta_0 = 100$ and $R=5$ as the radius of the interaction region. 
    The Roman numbers above the plot refer to the transition zones of
    Ref.~\protect\cite{Petit-Henon}.
    The horizontal lines under the graph of $T(h)$ identify the three 
    first-level blocks of Sec.~\protect\ref{sec-orbits} 
    and the three subblocks in one of them obtained at the second level 
    (see text).
    The gap between the two subblocks on the right of the large middle
    block is hardly visible.
(b) The blowup of zone IV.}
\label{fig-dwtime}
\end{figure}

In chaotic scattering processes, the dwell time function \cite{Eck-Jung}, 
measuring the time spent by a particle\footnote{%
As we have noted, the particle can be identified, for simplicity, 
with the smaller satellite if the other one is much larger.} 
in the region of strong interaction with the scatterer, shows irregular
behaviour similar to that of other scattering data.  
We can construct it by choosing a one-parameter family of initial
conditions and recording the dwell times for the trajectories.
In our case of satellite encounters, the family of initial circular orbits
parametrized by $\xi_0=h$ is a natural choice.
To give the initial conditions uniquely, we choose the origin of time so
that at $t=0$ the initial $\eta$ coordinate is a fixed value 
$\eta_0 \gg 1$, while (as a consequence of $h >0$ and $k=0$)
${\dot{\xi}}_0=0$ and ${\dot{\eta}}_0 < 0$ is obtained from $\Gamma$.
The incoming asymptotics then corresponds to moving downward along a
vertical straight line with 
$\xi_i (t)=h $ and $\eta_i (t)=\eta_0-|{\dot{\eta}}_0| t$.
We choose a circle with radius $R$ around the origin as the interaction
zone and define the dwell time $T$ as the time spent by the orbit inside
this circle.
The plot of the function $T(h)$ is shown in Fig.~\ref{fig-dwtime}a.

The irregular structure of $T(h)$ appears as a set of singularities 
crowding in four narrow regions separated by smooth behaviour.  
These regions themselves, called the transition zones in 
Ref.~\cite{Petit-Henon}, have a similar structure as shown by the 
blowup in Fig.~\ref{fig-dwtime}b.
Since the singularities correspond to trajectories asymptotically
trapped by the confined orbits of the chaotic set, the self-similar
pattern of singularities reflects the fractal structure of that set;
thus, an analysis of the set of singularities will provide important
information on the organization of the chaotic set \cite{BGO,Eck-Jung,KT}.
The fractal pattern of singularities suggests that we consider the
function $T(h)$ as built of a few basic blocks containing the
singularities and separated by smooth regions (valleys).  
In turn, these basic blocks are considered again as containing smaller
blocks separated by smaller regions of smooth behaviour.
This approach yields a whole hierarchy of blocks sitting on top of the set
of singularities: the basic blocks are on the first level of the
hierarchy, the subblocks contained in them go to the second level, etc.
The blocks on a given level of the hierarchy provide a coverage
of the set of singularities: the higher we look into the hierarchy,
the higher resolution we obtain in the coverage.

To carry out this hierarchic decomposition of the time delay function, we
need rules telling us how to break the continuous picture based on the
variable $T$ into the obvious discreteness introduced by the hierarchy.
These rules determining which of the singularities belong together in one
block at a certain level of the hierarchy should reflect the intrinsic
organization of the chaotic set.
As was shown in Ref.~\cite{KZLW} for a smooth potential model, this can be
achieved by linking the rules of block construction to the topological
complexity of scattering trajectories so that the orbits of the valleys
separating the blocks at a given level have the same degree of complexity
in their structures.
%

\begin{figure}
\vspace*{.5cm}
\centering\leavevmode \epsfysize=10cm\epsfbox{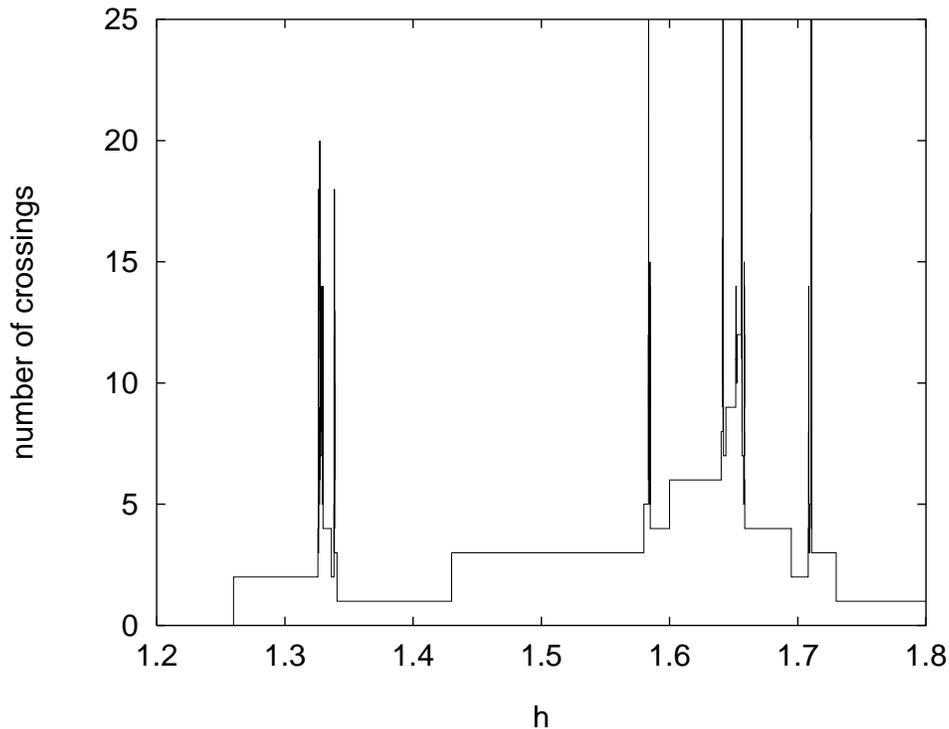} \\[.5cm]
\caption{%
The discrete dwell time function $N(h)$ for the same orbits as in
Fig.~\protect\ref{fig-dwtime}.}
\label{fig-ncross}
\end{figure}
One measure of the orbit complexity can be the number of crossings 
the orbits produce with a suitably chosen Poincar\'e section.
The function $N(h)$ giving the number of crossing of a trajectory from
our family of initial conditions with the surface $\eta =0$ can be
considered as a {\em discretized} dwell time function\footnote{%
In principle, we should record only crossings that
 happen in the same direction in a Poincar\'e section; 
 however, we just use the section to measure the scattering time, so
 this simplification will not cause any trouble.}:
it can be shown that all the periodic orbits of the system cut this
surface, so the longer a scattering trajectory follows a particular
periodic orbit, the larger the number of crossings $N$ become.

The discretized time delay function is plotted in Fig.~\ref{fig-ncross}; 
its similarity to Fig.\ref{fig-dwtime} is obvious. 
The discreteness introduced in the picture will make the decomposition 
process easier to implement, since the smooth changes in $T$ are now
replaced by plateaus of fixed heights that can be compared to one another.
In the next section, we will obtain the rules of block construction
for $N(h)$ [and thus for $T(h)$] by looking at the topological properties 
of scattering orbits in Hill's problem.

\section{Scattering orbits and the ternary hierarchy}
\label{sec-orbits}

\subsection{The topology of orbits}

The scattering orbits approach, along their incoming asymptotics, the
interacting region around the origin where they perform their central
parts consisting of localized and usually complicated motion (for shorter
or longer times, depending on the initial conditions), and finally escape
downward or upward.
Since it is the central part that can get close to the bounded orbits
of the chaotic set, we will concentrate on the central parts of
scattering orbits, ignoring differences in their escapes.  
In the hierarchic decomposition of $N(h)$ or any other scattering
function, we would like to set apart orbits from one another according
to the complexity of their central parts. 
For this purpose, the value $N$ is a good indicator but we also need 
a visual evaluation of the graph of the orbit to make necessary
distinctions between different orbit types. 
Our main goal is to represent the valleys of the dwell time function
with scattering orbits, so that the comparison of these orbits
can tell us whether two given valleys belong to the same level
in the hierarchy or not (we will say that a valley is of the 
$n$th level if the shortest block containing it belongs to level $n-1$).
For the comparison, we will identify some key elements in the orbit
plots so that the number of these elements in the orbit will give the
level number of the corresponding valley in the hierarchy.

\begin{figure}
\vspace*{.5cm}
\centering\leavevmode {\epsfysize=4.5cm\epsfbox{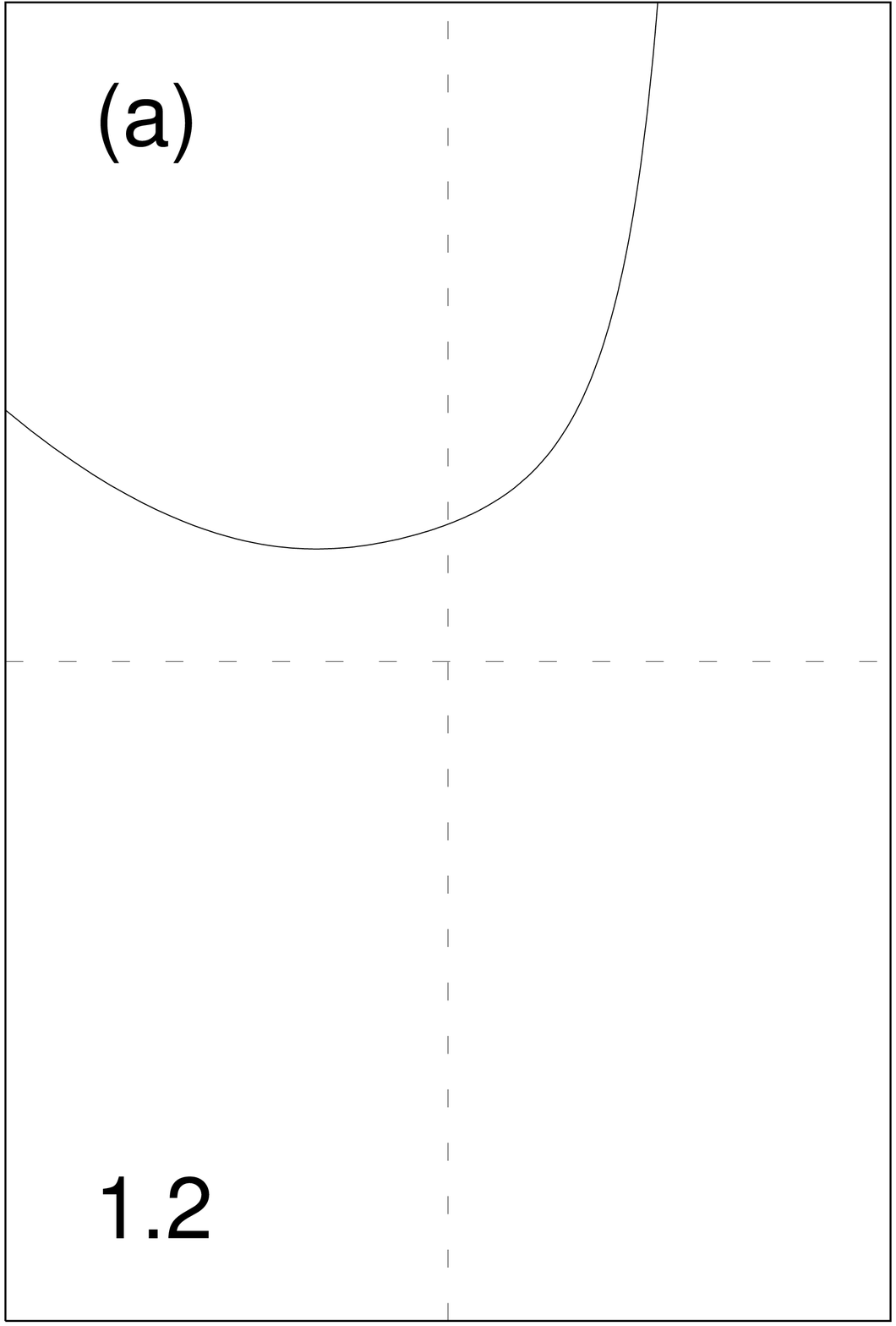}\hspace{.5cm}
                       \epsfysize=4.5cm\epsfbox{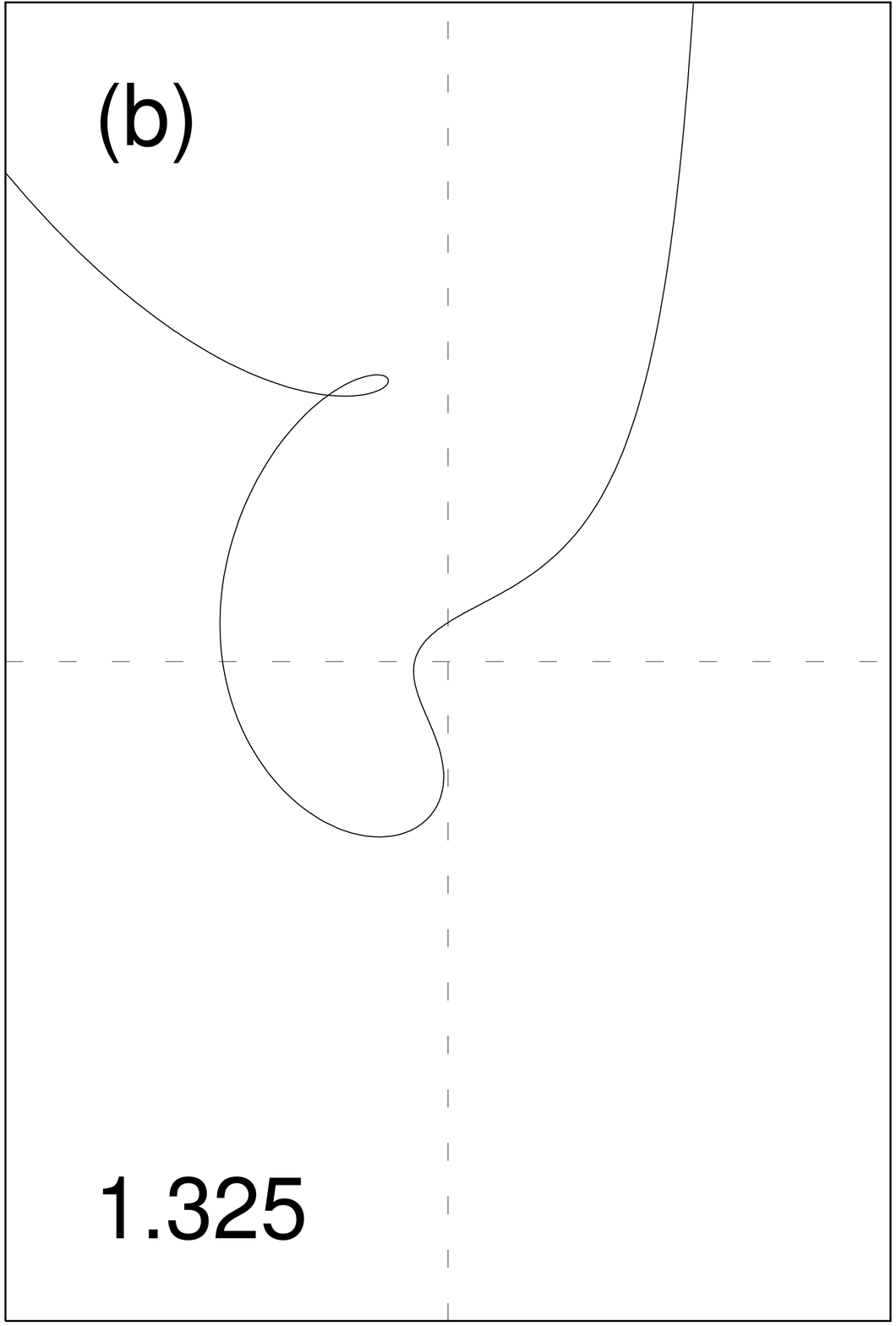}\hspace{.5cm}
                       \epsfysize=4.5cm\epsfbox{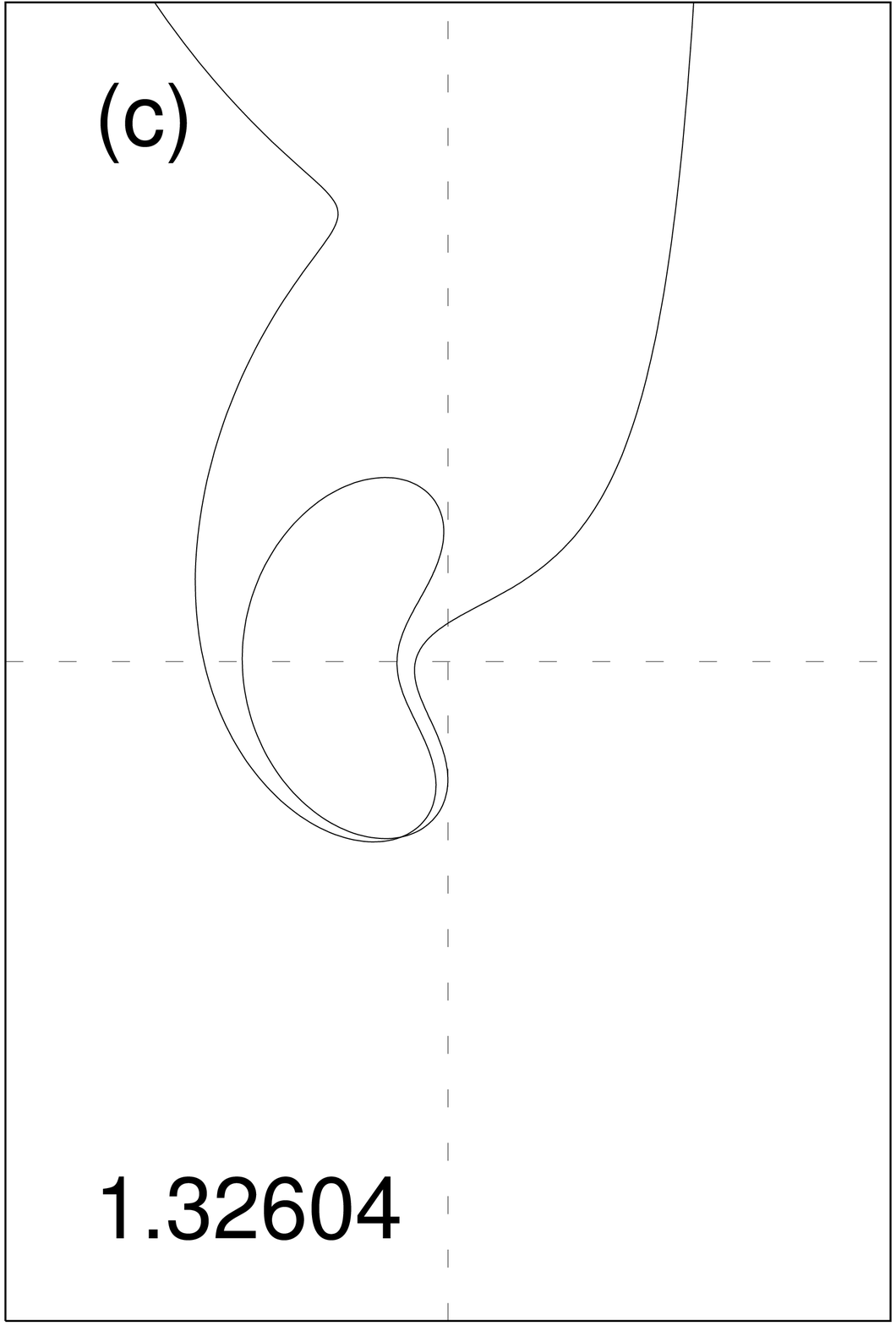}\hspace{.5cm}
                       \epsfysize=4.5cm\epsfbox{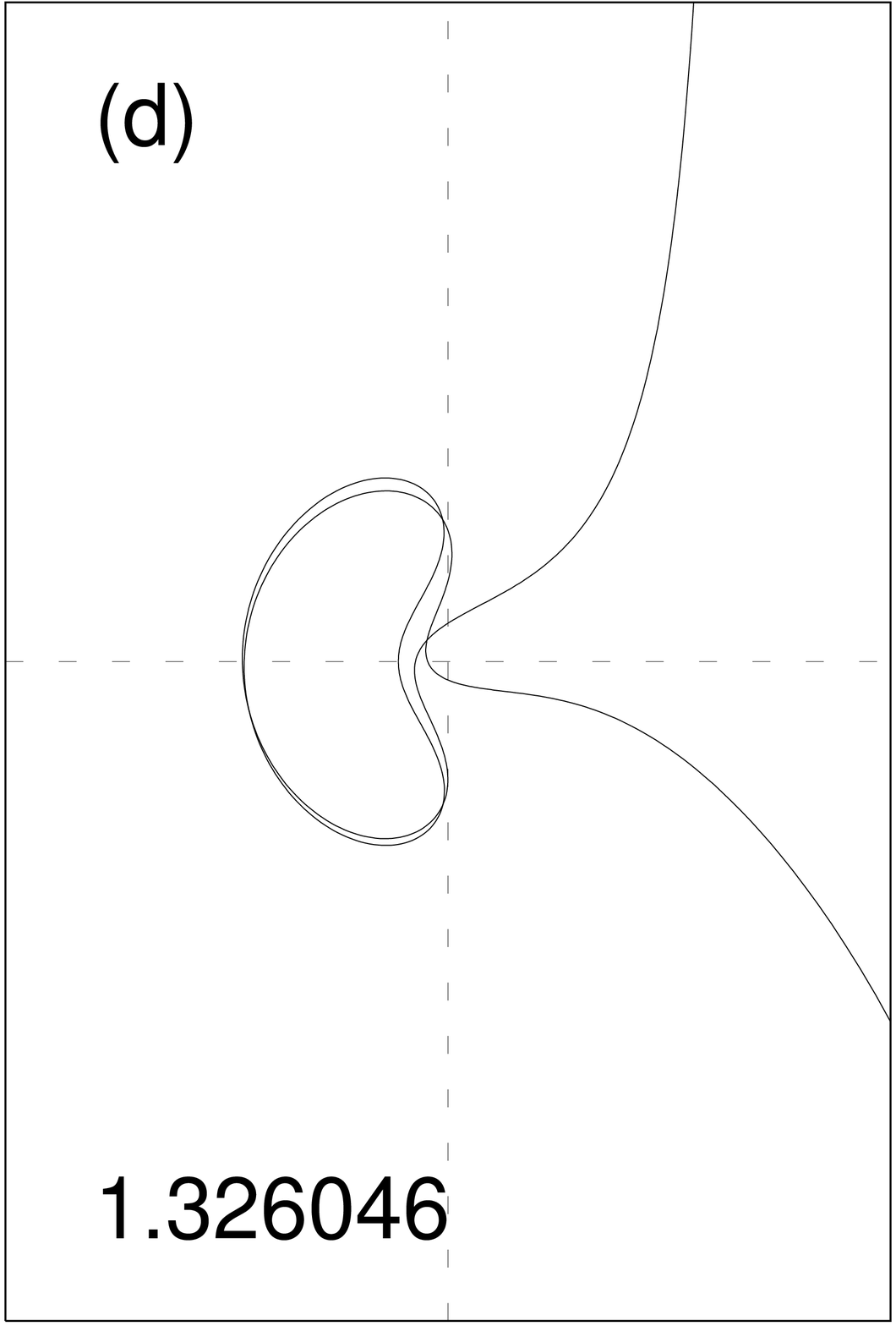}}  \\[.5cm]
\centering\leavevmode {\epsfysize=4.5cm\epsfbox{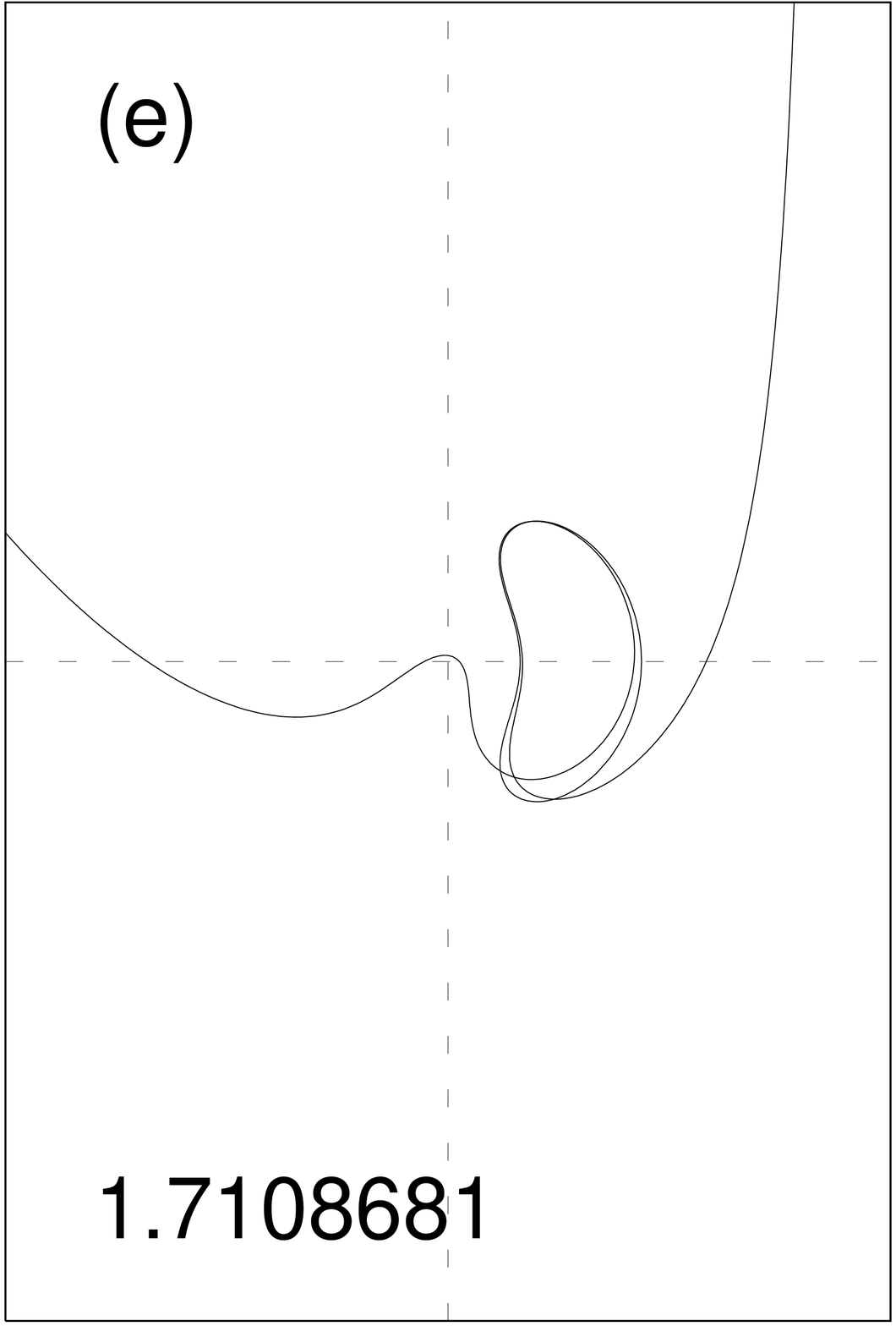}\hspace{.5cm}
                       \epsfysize=4.5cm\epsfbox{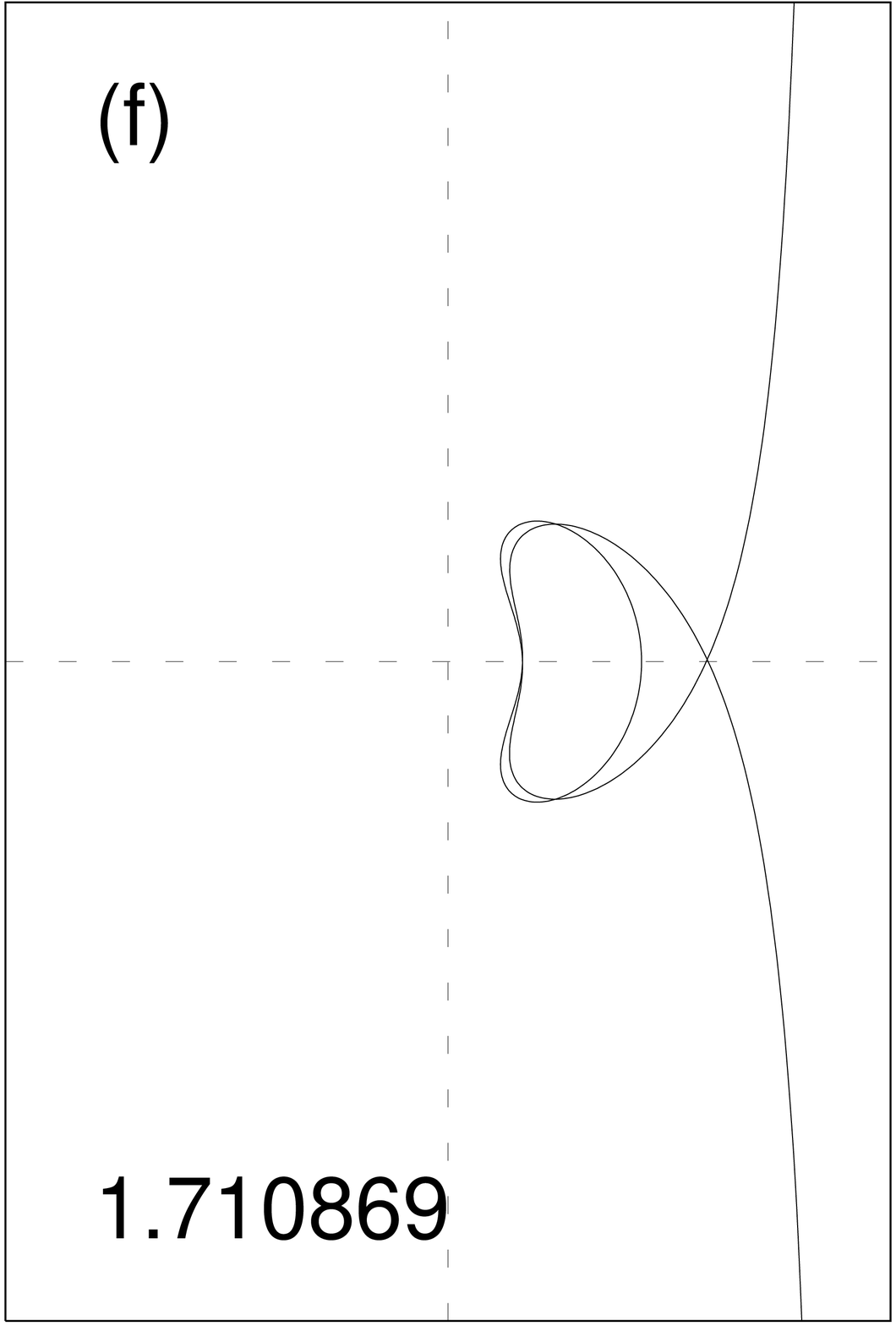}\hspace{.5cm}
                       \epsfysize=4.5cm\epsfbox{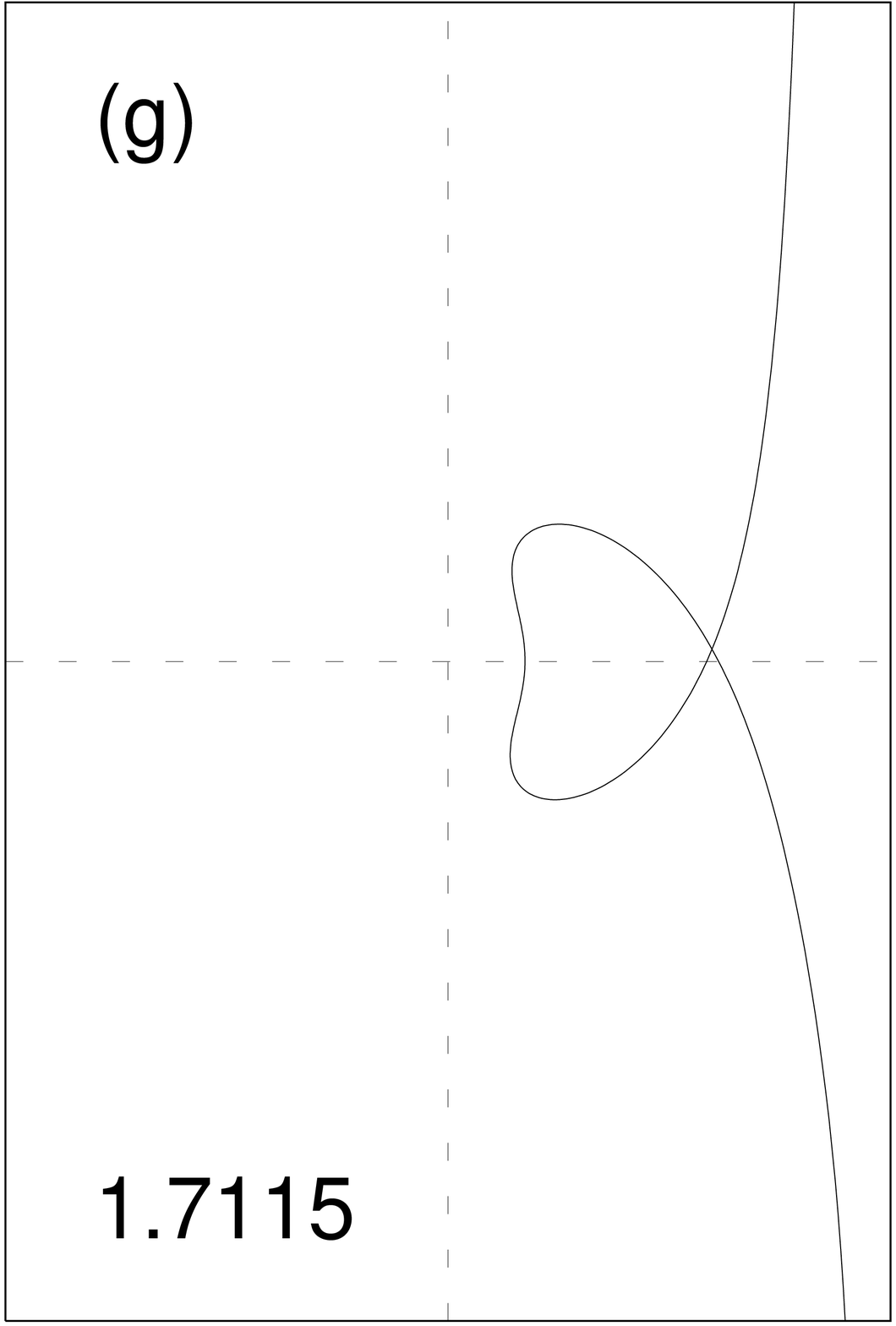}\hspace{.5cm}
                       \epsfysize=4.5cm\epsfbox{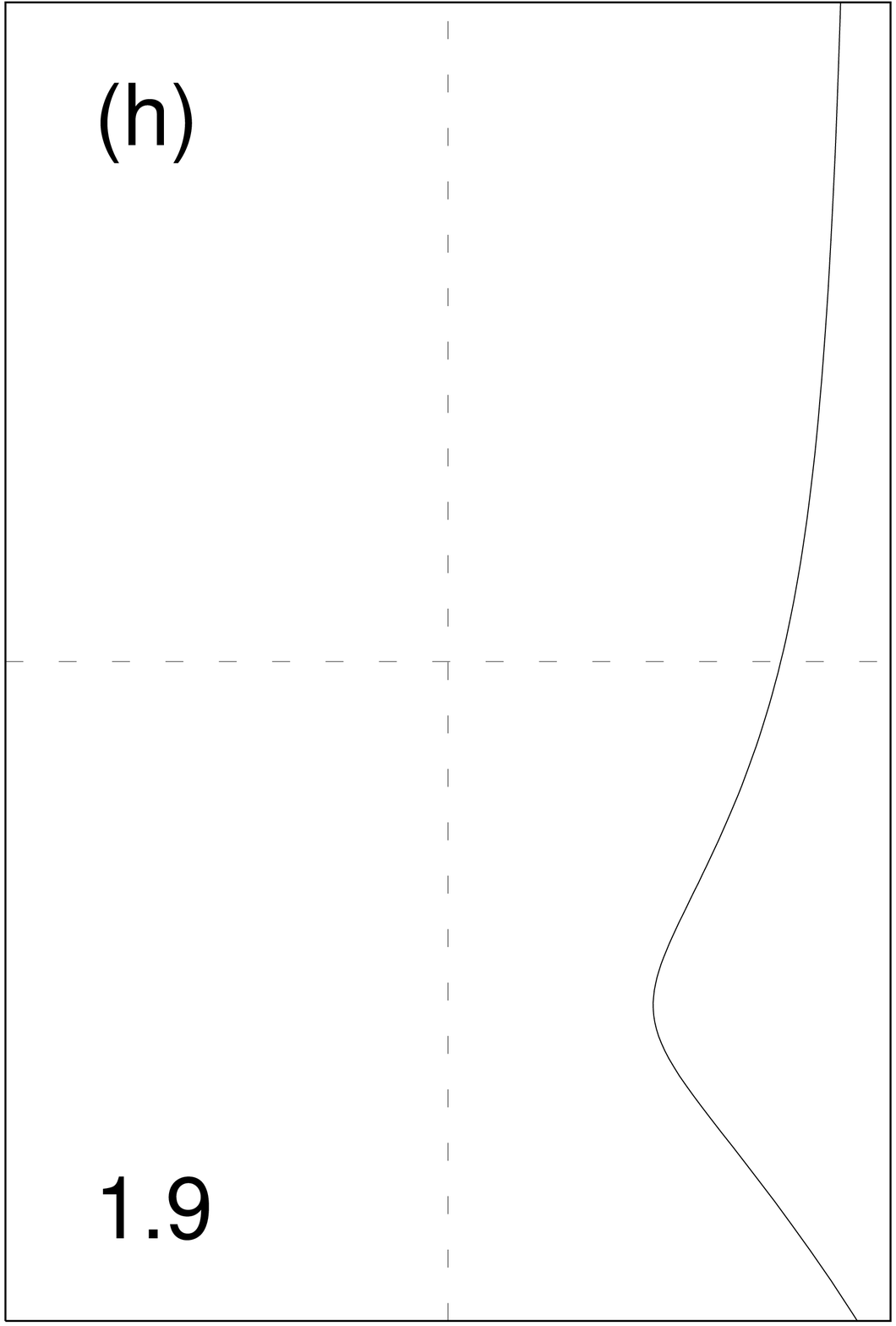}}  \\[.5cm]
\caption{%
The evolution of scattering orbits when approaching the outermost 
singularities of the dwell time function.
All the windows show $\xi$ horizontally and $\eta$ vertically in the
ranges $-2 < \xi < 2$ and $-4 < \eta < 4$ (with the axes as dashed lines).
The numbers on the plots give the corresponding values of $h$ in the
initial conditions. 
The orbits in Figs.~\protect\ref{fig-traj1}d and e are from 
inside the irregular domain.}
\label{fig-traj1}
\end{figure}
As can be seen from Figs.~\ref{fig-dwtime} and \ref{fig-ncross}, all the
singularities of the dwell time function are contained in the domain
between $h_a \approx 1.326$ and $h_b \approx 1.711$
\cite{remark}.
Above $h_b$, all the trajectories escape downwards, the simplest ones
making only one crossing with the $\xi$ axis, on the positive 
$\xi$ side (Fig.~\ref{fig-traj1}h).
Approaching $h_b$ from above, the orbits make more and more turns around 
a kidney-shaped periodic orbit (Figs.~\ref{fig-traj1}g and f), developed 
from the Lagrangean point $L_2$ of the system \cite{Petit-Henon}, 
before the escape.
These turns appear in $N(h)$ as a sequence of shortening plateaus 
increasing in height in steps of two [while there is a continous
divergence in $T(h)$]; the shortening is geometric governed by the stable 
eigenvalue of the periodic orbit.
On the other side of $h_b$, the escape may also happen upwards
(Fig.~\ref{fig-traj1}e).
A similar phenomenon can be observed when approaching $h_a$ from
below (Figs.~\ref{fig-traj1}a--d): 
all the orbits escape upwards (the simplest ones with $N=0$),
while winding more and more around a similar periodic orbit, and
for $h > h_a$ escapes can appear in both directions.

Looking at the larger valleys of $N(h)$ inside the irregular domain, 
we can observe that each valley is dominated by two plateaus with a 
difference of 2 in their $N$ values; they also lie lowest within the valley.
These plateaus correspond to the simplest possible orbits in the valley, 
their difference in $N$ coming from the fact that the escaping part
may cross (twice) the $\xi$ axis.
Approaching the edges of the valley, $N$ diverges, in the same way as
observed outside the irregular domain, when the trajectories get closer 
and closer to one of the kidney-shaped orbits but still escape in the same 
direction.
Thus the only significant difference in the structure of two orbits
from the same valley is in the number of turns around the same kidney-cycle
immediately before escaping \cite{gate}.
If we consider these turns as belonging to the escape process, 
then we can claim that central parts of all the orbits from a given 
valley have the same topological structure.
In contrast, orbits taken from different valleys have characteristically
different central parts.
This property makes it possible to represent a valley with one of its 
scattering orbits; we will choose this from the widest plateau with a height
of two above the minimum of the valley.
Figure~\ref{fig-traj2} shows scattering trajectories representing the
largest valleys and some of the narrower ones.

\subsection{The ternary hierarchy}

Now we are in a position to start the decomposition procedure.  
The blocks of a given level can be obtained by removing the
appropriate valleys from the blocks of the previous level.
The inclusion of the final kidney-turns in the escaping part puts
the border between a valley and the neighbouring block right to 
a singularity.
To decide whether a certain smooth region between two singularities is
to be deleted at a given level, we analyze the structure of its 
representant scattering orbit: the valleys with the simplest trajectories 
within the given block will be removed to yield the new subblocks.
The irregular region of $N(h)$ can be separated into first-level blocks
by comparing the trajectories representing the largest visible smooth
regions between the transition zones  of Ref.~\cite{Petit-Henon}.

\begin{figure}
\vspace*{.5cm}
\centering\leavevmode {\epsfysize=4.5cm\epsfbox{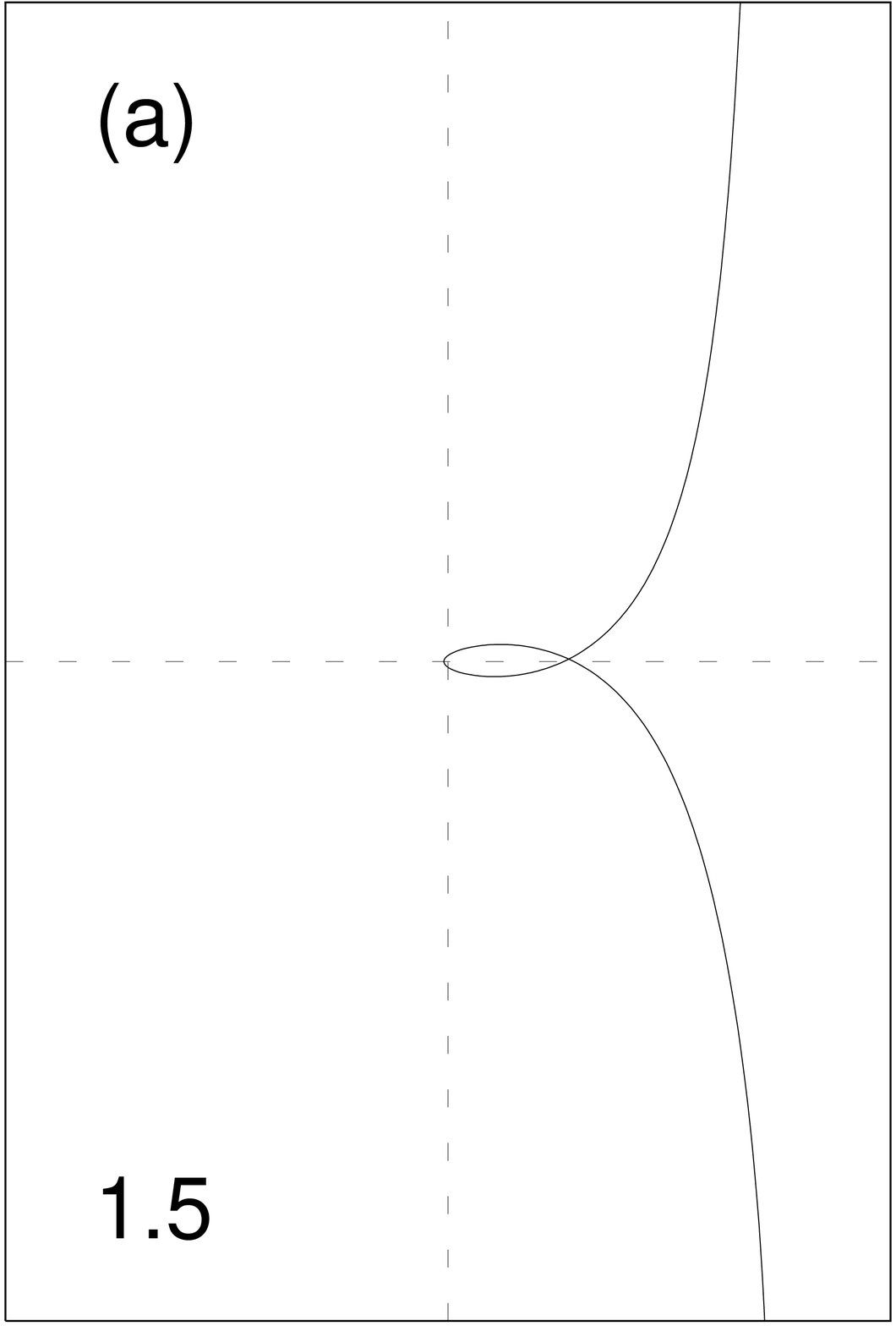}\hspace{.5cm}
                       \epsfysize=4.5cm\epsfbox{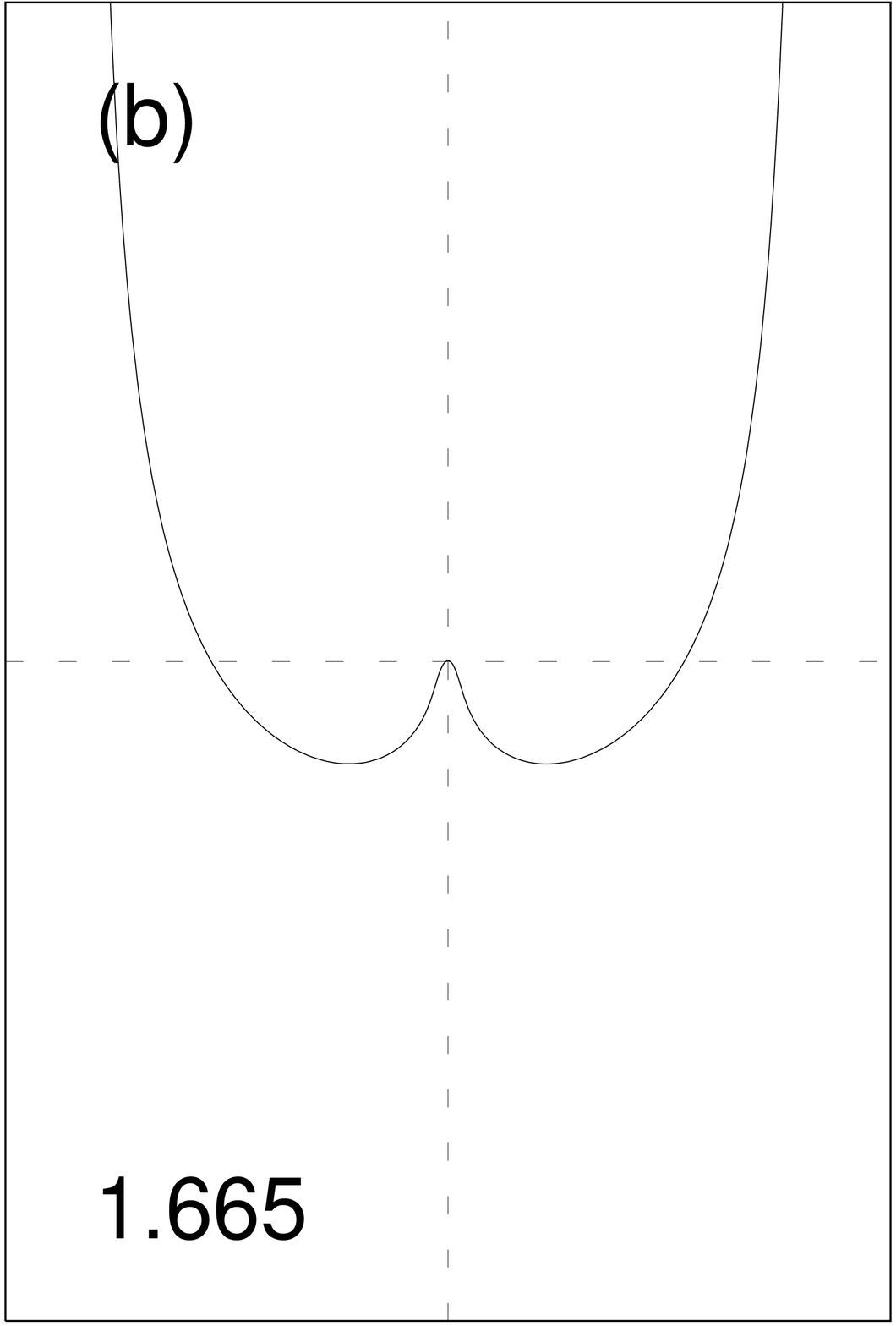}\hspace{.5cm}
                       \epsfysize=4.5cm\epsfbox{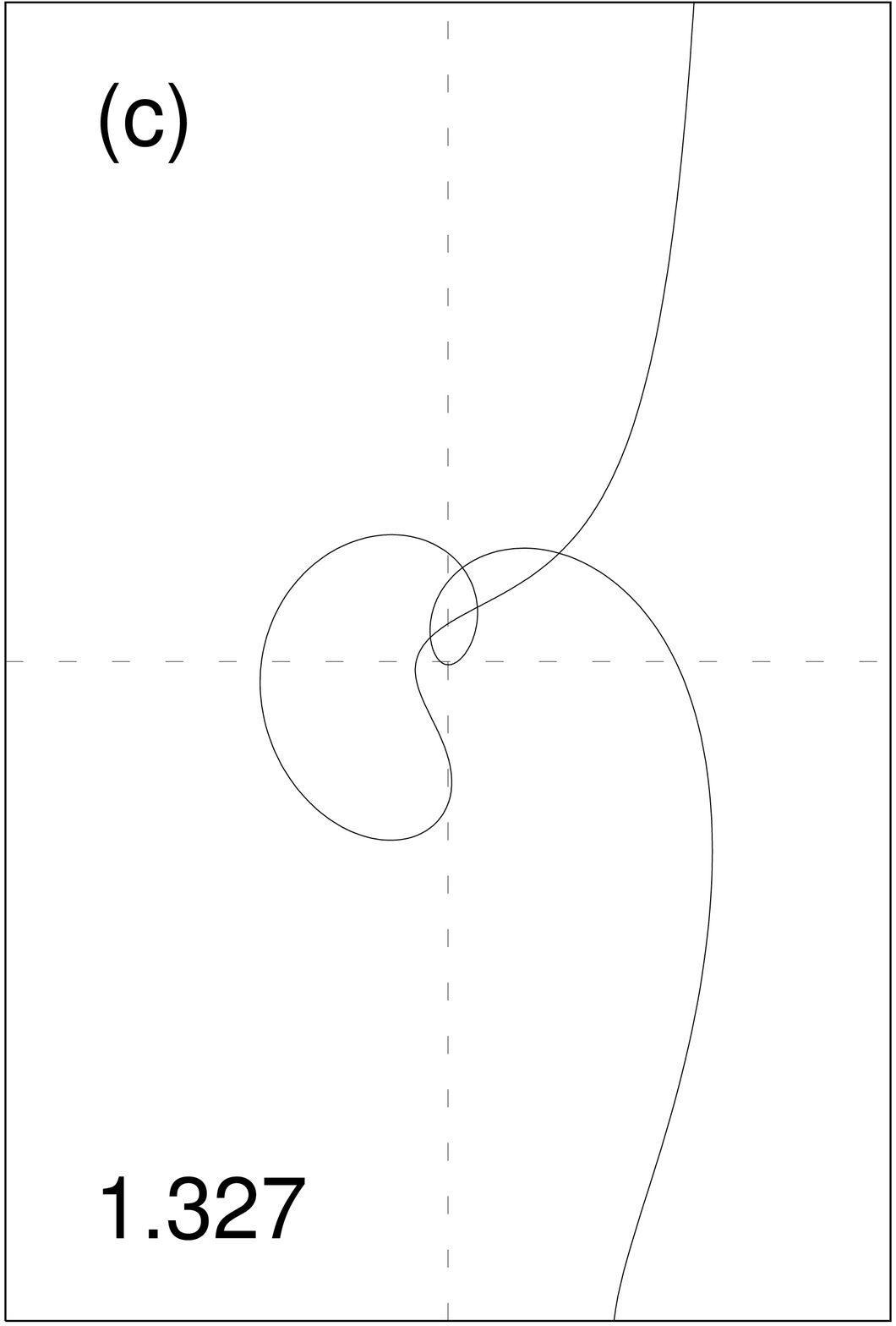}\hspace{.5cm}
                       \epsfysize=4.5cm\epsfbox{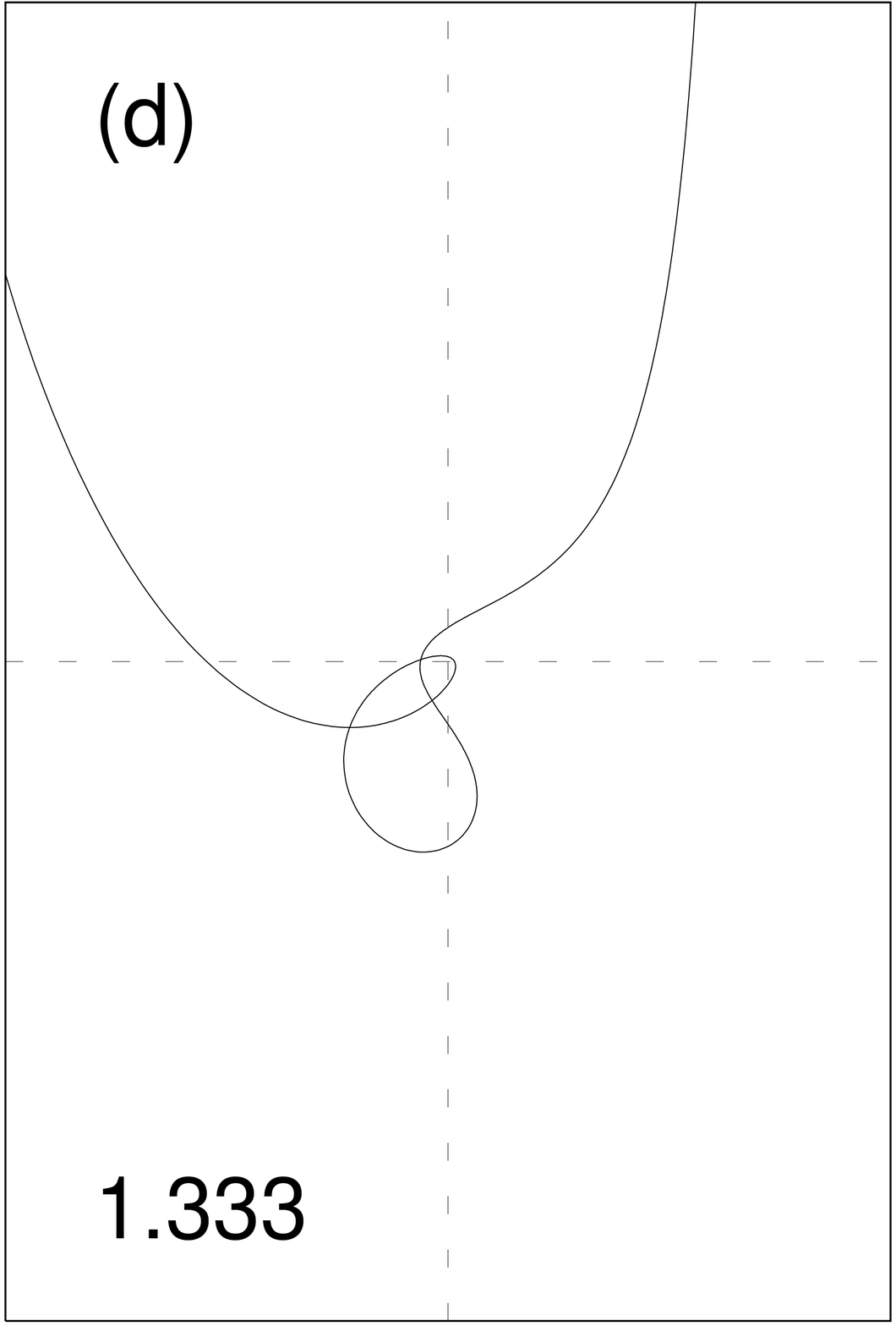}}  \\[.5cm]
\centering\leavevmode {\epsfysize=4.5cm\epsfbox{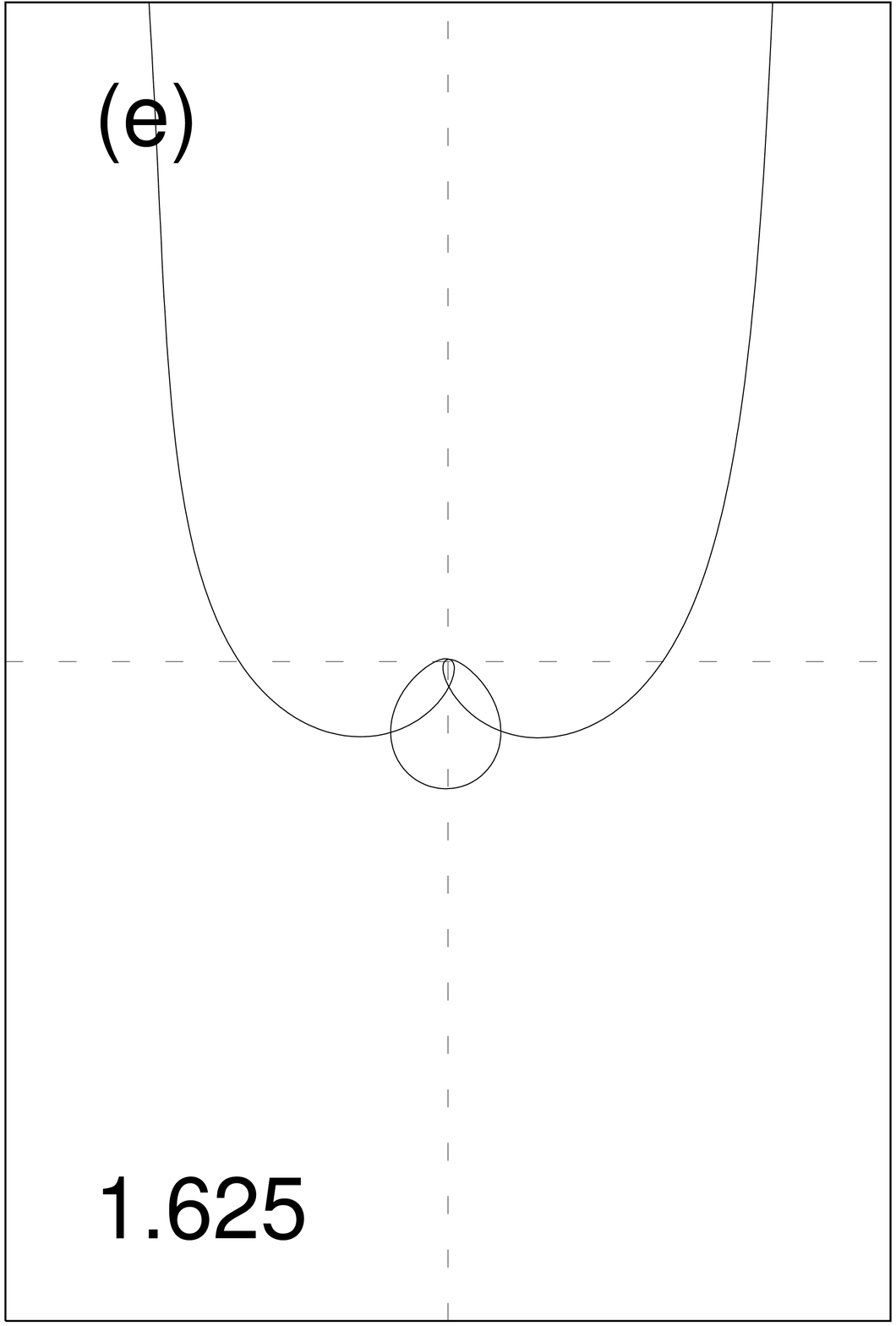}\hspace{.5cm}
                       \epsfysize=4.5cm\epsfbox{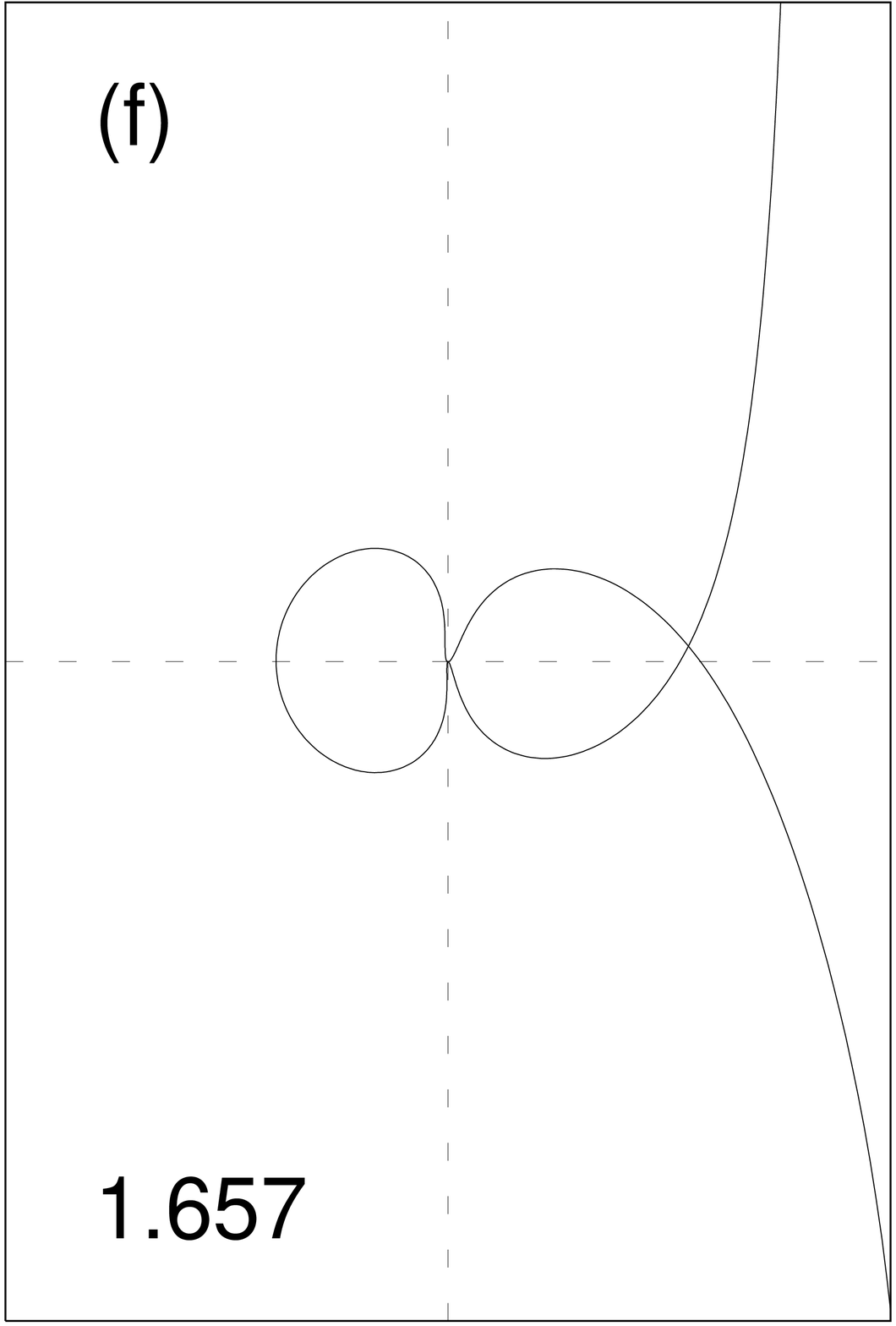}\hspace{.5cm}
                       \epsfysize=4.5cm\epsfbox{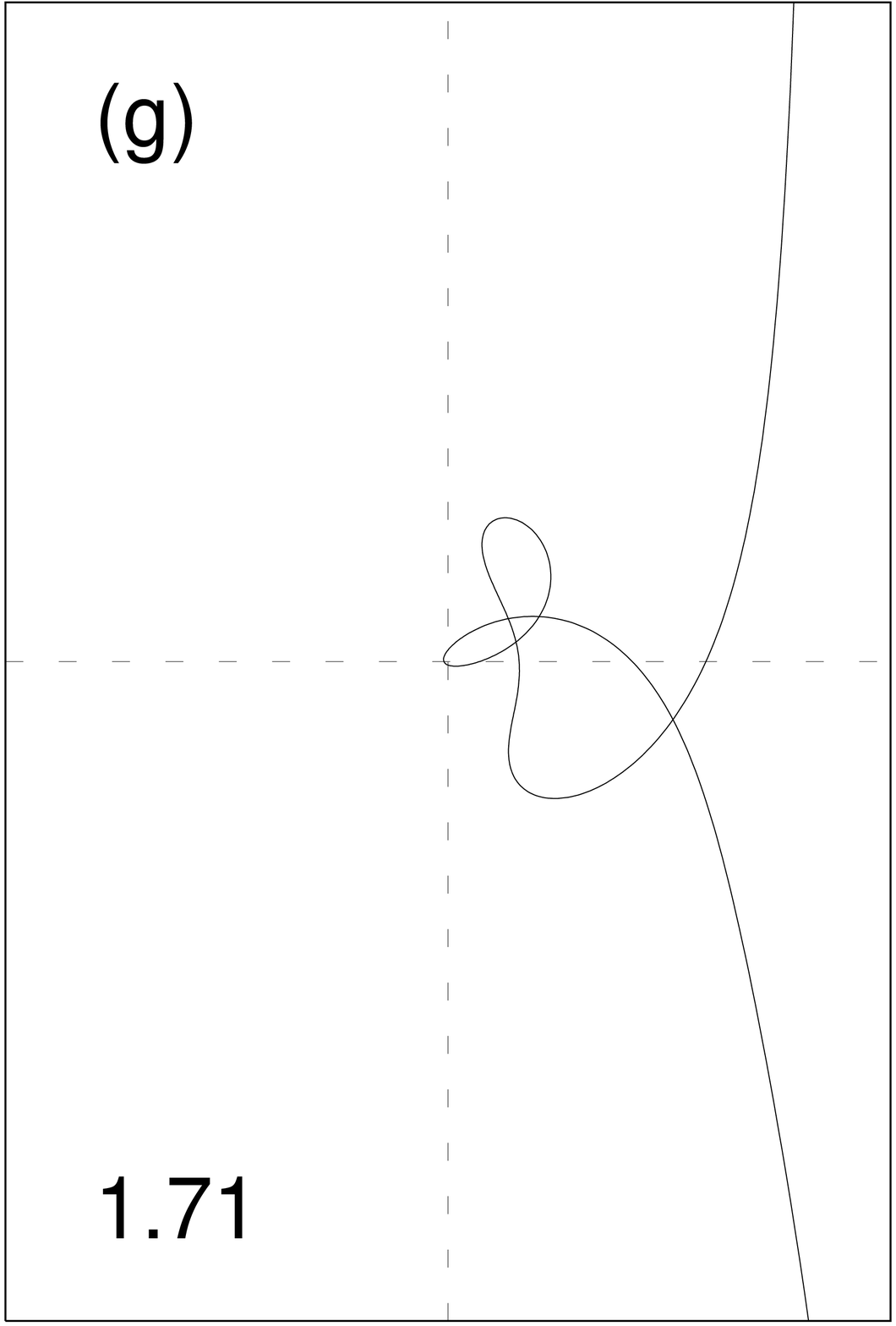}\hspace{.5cm}
                       \epsfysize=4.5cm\epsfbox{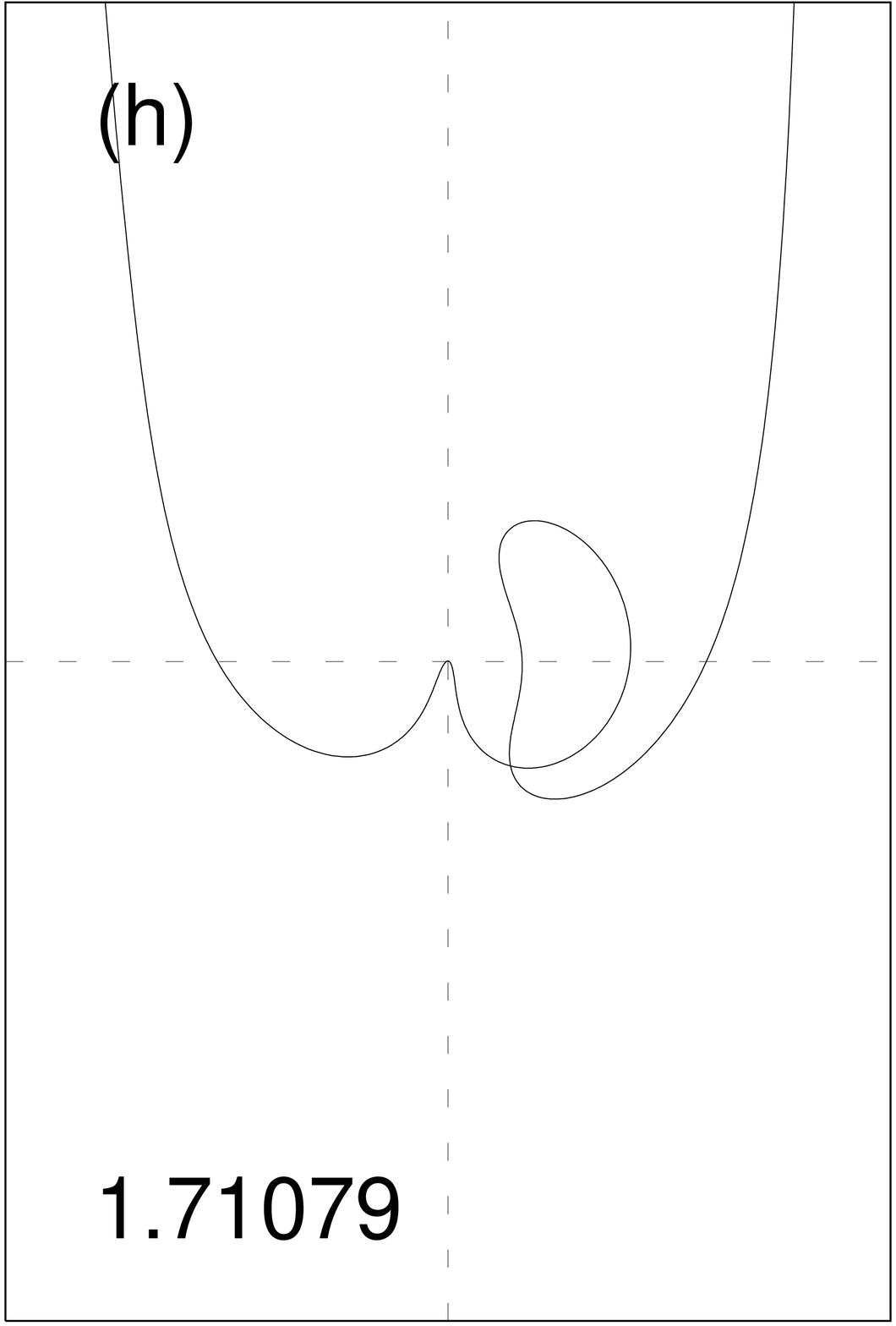}}  \\[.5cm]
\caption{%
The scattering orbits that represent the most prominent valleys defining 
the blocks on the first and second levels.}
\label{fig-traj2}
\end{figure}

The three orbits shown in Figs.~\ref{fig-traj2}a, e, and b sit in the
valleys separating zones IV--III--II--I, respectively.  
We may immediately conclude that this separation into four blocks 
does not satisfy the ``equal complexity'' requirement:
the orbits of Fig.~\ref{fig-traj2}a and b are clearly of the same type and
complexity (the central part is just one close approach of the
origin), but the orbit in Fig.~\ref{fig-traj2}e is obviously more
complicated performing two close approaches of the origin.
Therefore, as a second attempt, we consider this middle valley as 
separating blocks on a higher level and consequently, zones II and III 
as parts of one larger first-level block ranging from about 1.58 to 1.66;
the regions I and IV will form the two other blocks in the first level 
of the hierarchy (Fig.~\ref{fig-dwtime}a).

The fact that we obtained three blocks in the first step suggests that 
we should look for a ternary hierarchy dividing each block into three 
at the next level.
Indeed, these blocks show, when magnified, an inner structure similar
to the whole irregular region: there are two simplest trajectory types 
within each block, plotted in Fig.~\ref{fig-traj2}c--h. 
The topological similarity between the central parts of 
Figs.~\ref{fig-traj2}e and f is now obvious (two close approaches to 
the origin), while in the other orbits the first close approach is
replaced by a prograde bend, essentially a part of the kidney-shaped orbits. 
Considering the close approaches and the prograde bends as basic
structural elements of the central parts of trajectories, the first-level
blocks contain orbits that start their central parts with one of these
elements, the same element within one block. 
Similarly, the blocks on the second level gather together orbits starting
their central parts with the same pair of basic elements, a different
pair for each block.
In turn, the magnification of the second-level blocks produces singularity
structures resembling to the picture of Fig.~\ref{fig-ncross} again, 
and all the 18 trajectories representing the dividing valleys can be 
decomposed into the three basic elements mentioned.

We may conclude that in general all the orbits from an $n$th-level block 
share the first $n$ elements of their central parts, and the shortest
of these orbits, with just $n+1$ elements, sit in the valleys that will 
split the block into subblocks at the next level.
Thus, when analyzing a scattering orbit from a given valley, 
the number of these elements in the central part gives the level number 
of the valley and consequently the (largest) blocks it separates.
The division rules can easily be formulated in terms of the discrete dwell
times too: given the minimal $N$ values for the two
valleys next to a particular block, a simple rule will give the minima
of the two valleys splitting that block into three at the next level.
If the block intended to split sits in one side of a larger
(previous-level) block next to a valley (outside the larger block) 
with a minimum $N_0$, then we look for valleys with minima $N_0+1$ and
$N_0+2$ inside; however, if the block sits in the middle with neighbouring
valley minima $N_1$ and $N_2$, then the new valleys are those with minima
$N_1+3$ and $N_2+3$.

\subsection{The stable island}

\begin{figure}
\vspace*{.5cm}
\centering\leavevmode \epsfysize=10cm\epsfbox{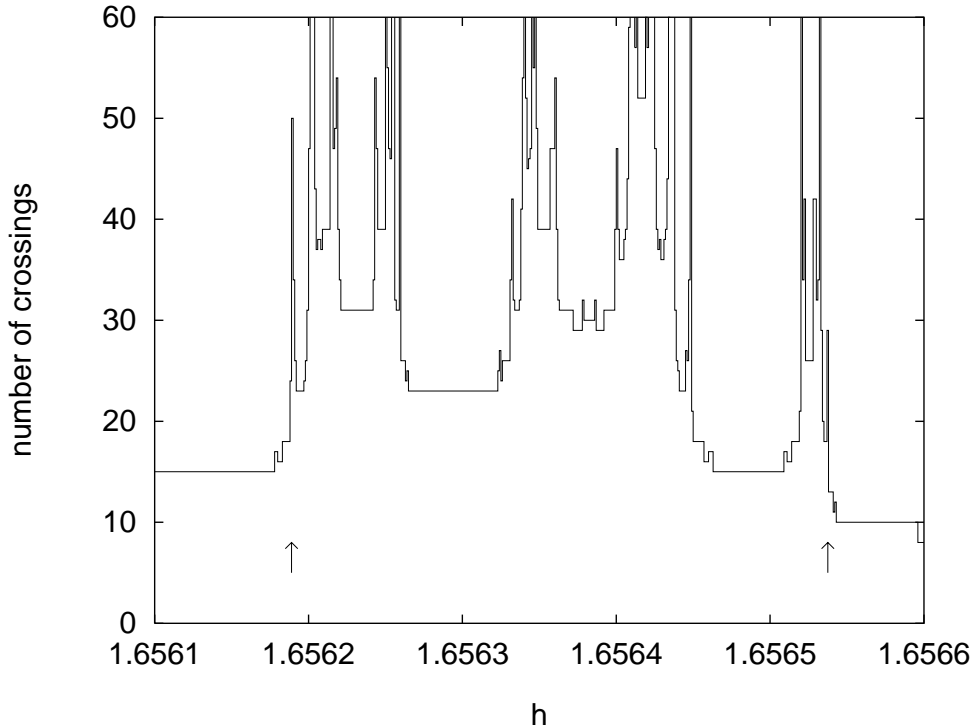} \\[.5cm]
\caption{%
The magnification of the fifth-level block with an anomalous inner structure. 
Notice the difference between this plot and Fig.~\protect\ref{fig-ncross}.
The arrows at values $h_{3a}$ and $h_{3b}$ mark the positions of
the two scattering orbits running directly into the 3-cycle.}
\label{fig-badblock}
\end{figure}

The ternary organization established by the rules above can be followed 
climbing higher in the hierarchy, and we found that it is complete up to 
the fifth level.
In constructing the sixth level of the hierarchy, the rules of the
decomposition cannot be applied to split one block in the center of
the level since it has an inner structure different from that of the 
other blocks (Fig.~\ref{fig-badblock}).
This particular block contains trajectories that are built of only close
approaches to the origin avoiding the kidney-shaped orbits, and
are organized in a way which is not captured by the rules we used before. 
\pagebreak

\begin{figure}
\vspace{.5cm}
\centering\leavevmode {\epsfysize=5.5cm\epsfbox{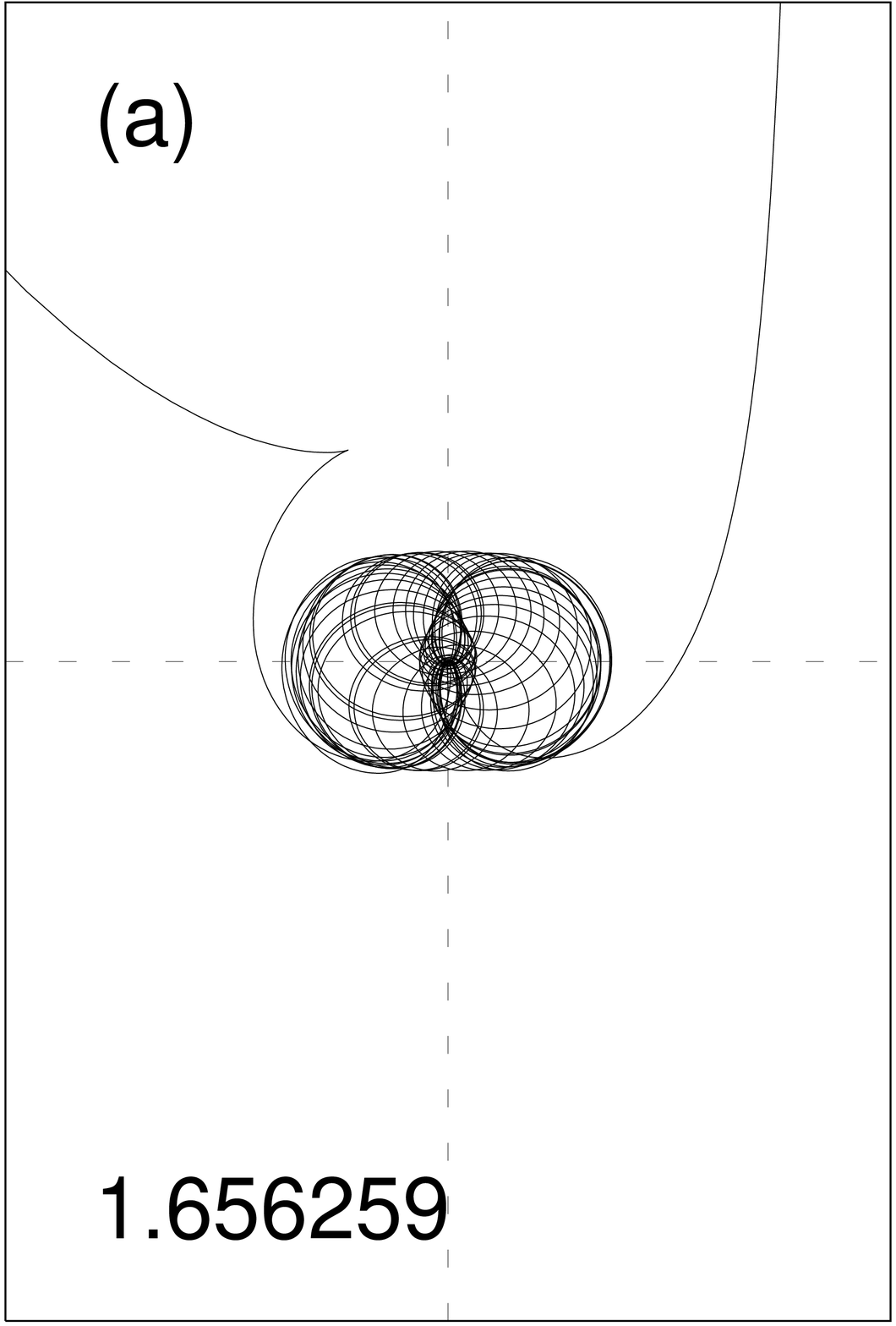}\hspace{1cm}
                       \epsfysize=5.5cm\epsfbox{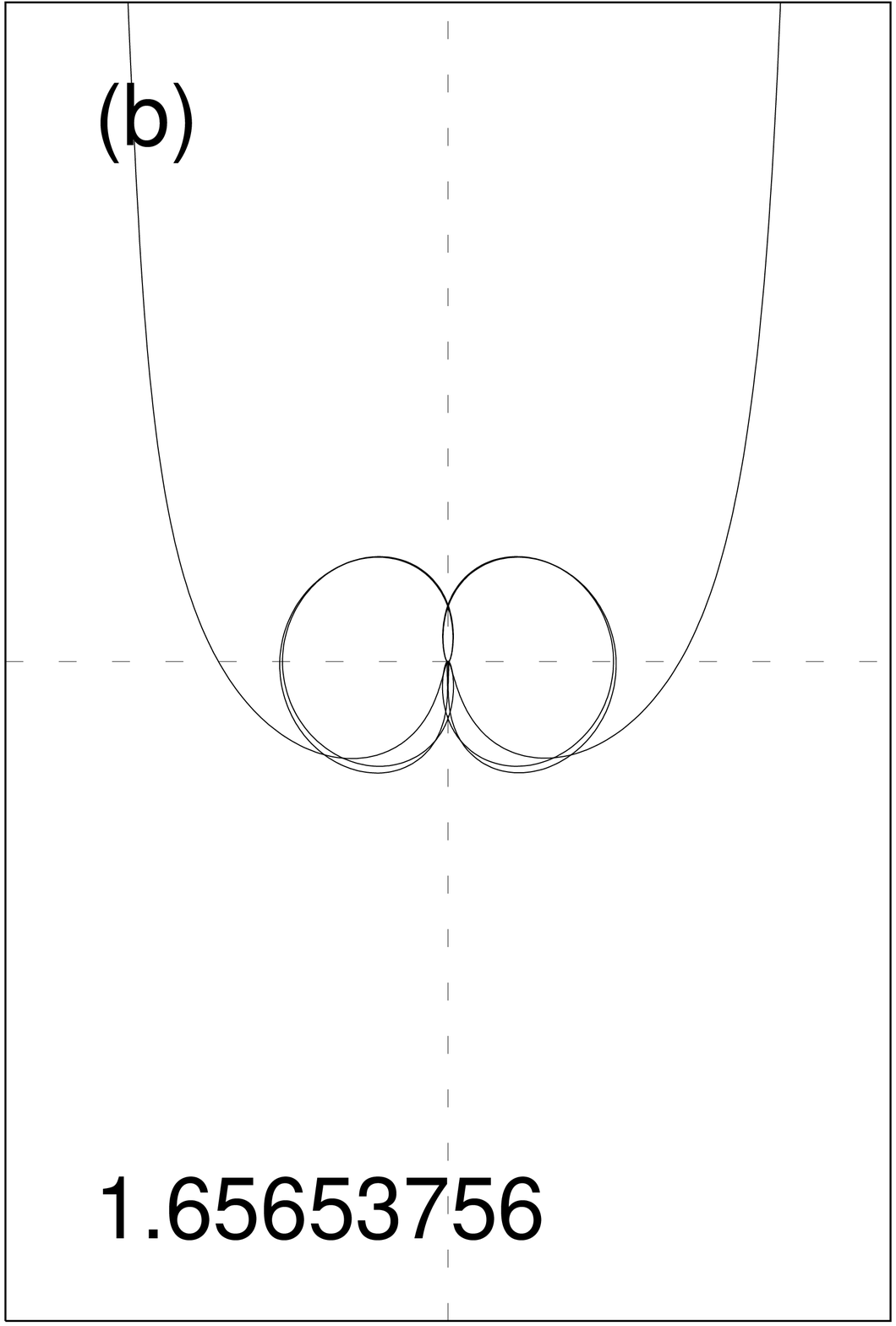}} \\[1cm]
\caption{%
(a) One of the scattering orbits from inside the stable block in 
Fig.~\protect\ref{fig-badblock} with anomalously large dwell times
($N=108$ for this one). 
(b) A scattering trajectory with $h \approx h_{3b}$ 
approaching the period-3 cycle in the vicinity of the stable island.}
\label{fig-traj3}
\end{figure}

The reason for this anomaly is that these orbits are close to a region
of {\em stable motion}
corresponding to a bound configuration of the two satellites. 
In the Poincar\'e section, the stable motions form a KAM island with the
well-known hierarchy of periodic orbits with chains of smaller islands
around the edge, and the rules of organization for that hierarchy
differ from those of the ternary organization. 
Although the scattering trajectories cannot reach the inside of the
island, the complicated structure around the edge
of the island results in scattering trajectories that are trapped for
anomalously long times in its vicinity (Fig.~\ref{fig-traj3}a).
The existence of these anomalous scattering orbits was noted in
Ref.~\cite{Petit-Henon}; now we can explain it by the presence of the
island.

The image of the primary ``stable'' block reappears in the middle of 
other ``ordinary'' blocks at higher levels of the hierarchy copied there 
by the dynamics.
In general, keeping away from the two kidney-shaped orbits for a certain
number of steps in the refinement process of any block in the
hierarchy will reveal these island-approaching orbit blocks with a
non-ternary inner structure.
On the other hand, the subtleness of these flaws in the rules describing
the hierarchy also means that the ternary organization remains a good 
approximation to the hierarchic properties everywhere in the chaotic set 
except in a narrow region around the island.

It is important to note that near the two edges of the primary stable
block there are two values $h_{3a}$ and $h_{3b}$ of the initial condition 
for which the scattering trajectory runs directly into an eight-shaped 
hyperbolic orbit with $N=6$ and two close approaches of the origin, 
avoiding the kidney-cycles (Fig.~\ref{fig-traj3}b).
In a proper Poincar\'e section, where we record only crossings with
e.g.\ $\dot{\eta} > 0$, this orbit appears as a period-3 cycle.
In fact, the close approaches as basic elements of orbit structure
connected to the rules of the ternary hierarchy reflect encounters
with this cycle.
In other words, the structure of the family of scattering orbits is
organized not only by the two kidney-cycles, as claimed in
Ref.~\cite{Petit-Henon}, but also by this 3-cycle representing the island.  
This also means that the chaotic set itself is built on these
three orbits as its pillars.

\pagebreak
The 3-cycle lies very close to the island, and its stable and unstable
manifolds form a ``cage'' around it: the orbits that enter the cage form
the irregular inner structure of the stable blocks.
The presence of the cage also ``hides'' the island for scattering 
trajectories as long as they stay away from its immediate vicinity:
the irregularities caused by the island in the hierarchy are covered by the 
hyperbolic behaviour of the neighbouring 3-cycle.
The closeness of the period-3 orbit to the island is a consequence of
the  delicate interplay known as {\em squeezing} between a
stable island around a fixed point and a nearby period-3 resonance
\cite{vdWCVP} found in generic Hamiltonian systems with a quadratic
nonlinearity.

\section{Discussion}
\label{sec-discuss}

We have shown that the singularity structure appearing in the dwell
time function in Hill's problem can be decomposed into blocks ordered 
in a ternary hierarchy which remains a good approximation even in the
presence of the distorting effects of a stable island.
The main tool of the decomposition is the analysis of the topological
structure of scattering trajectories: it can provide, in
conjunction with the discretized dwell time function $N(h)$,
the information we need on the hierarchic properties. 
In the simplest examples of chaotic scattering like the three-disk
model, the discretized dwell time function in itself can reveal the 
hierarchic structure immediately \cite{KT}: 
cutting its graph at a height $n$ will yield all the blocks in the
$n$th level.
However, in more complicated cases like the model discussed here, 
the structure of the graph is less revealing, so we need the extra 
information embodied in the scattering orbits.
The orbit plots we used in our analysis were mostly the same as
produced in Ref.~\cite{Petit-Henon}, but we surveyed them systematically to 
uncover the details of the hierarchy.

This hierachic analysis is not restricted to the dwell time function:
it can be carried out on any scattering function  (e.g. $h'(h)$ used
in Ref.~\cite{Petit-Henon}) that mirrors in
its singularities the structure of the chaotic set.
We have chosen the dwell time function $T(h)$ because of its close
relationship with $N(h)$.
The analysis of Hill's problem reported here also provides 
another example illustrating the capabilities of the hierarchic
analysis in models of chaotic scattering with two degrees of freedom.

The hierarchic information can be used in calculations of various
global quantities characterizing the scattering process. 
The main tool for such calculations is the thermodynamic formalism 
\cite{thermo}, where the lengths $l_i ^{(n)}$ of blocks at level $n$
of the hierarchy are put into the scaling sum \cite{KT}
\begin{equation}
      \sum_i {(l_i^{(n)})}^{\beta} \sim e^{-\beta F(\beta)n}          
\end{equation}
to produce the free energy function $F(\beta)$. 
Specific values of this function yield a few important scaling
quantities \cite{KT}.
The most notable ones are: 
(i) the topological entropy  $K_0 = - \beta F(\beta)|_{\beta =0}$ 
characterizing the growth rate of the numbers of blocks from one 
level to the next ($K_0 = \log 3$ for a ternary hierarchy), 
(ii) the fractal dimension $D_0$ of the set of singularities [from
$F(D_0)=0$], and (iii) the escape rate $\kappa = F(1)$
describing the exponential decay of survival of scattering
trajectories within the interacting region.
The number $\kappa$ also describes the exponential shrinking of the
total length of blocks at a given level if we go higher and higher in
the hierarchy.
In principle, the nonhyperbolic contributions due to the island produce 
nonexponential scalings, leading to $D_0=1$ and $\kappa=0$, but the
screening effect of the hyperbolic 3-cycle usually ensures that the 
scaling remains exponential for levels below a certain crossover value.

The hierarchic structure can be exploited to obtain reliable results for
integrals like $  I = \int\limits_0^{\infty} h \Delta (h^2) \, dh $
(with $\Delta (h^2) = h'^2 - h^2$) describing the rate of energy
transfer in a planetary ring system composed of many particles of
equal mass \cite{Petit-Henon}.
Because of the irregular behaviour of the scattering function $h'(h)$
the integral cannot be determined very accurately by a usual 
algorithm---a fact that has already been noted in Ref.~\cite{Petit-Henon}.
However, we can use the hierarchic information to make the
calculations more consistent \cite{KZLW}: we can evaluate the
integral within the smooth valleys separately and add up their
contributions in the order they appear in the hierarchy, starting from
the lowest level. 
After the first $n$ levels, the regions still missing from the integral 
are the blocks at level $n+1$ and higher.
Since their total length decays as $\exp(-\kappa n)$, our procedure
converges exponentially fast (as long as we can neglect the
nonhyperbolic effects associated with the island).

Concerning the validity of our findings, the particular choice of 
circular initial orbits ($k=0$), used in Ref.~\cite{Petit-Henon} and in 
our study too, is not special in the sense that a wide choice of the 
possible combinations of the parameters in the general form of the 
asymptotic motion can lead to similar results.
The only requirement is that the line representing the initial conditions 
in phase space cut the stable manifolds of the chaotic set \cite{KT}.
As we have seen, the ternary structure found in our case rests on 
the two kidney-shaped orbits and the period-3 cycle near the stable
island of bound motion.
These basic orbits can be found in Hill's problem with the same
properties for a certain range of $\Gamma$ values, 
so if our set of initial conditions remains within that
range, then we find the same organization (but, of course, with
different scaling ratios).

Finally, it is worth noting that the presence of the stable island remains
the main difference between Hill's problem and the inclined billiard
model \cite{Henon-bill}, proposed as a simple alternative
with similar properties, where a ball bounces on two large 
overlapping disks in a gravitational field.
The hierarchic organization found there was simply binary 
built around the two fixed points of vertical bouncing 
(analogous to the kidney-cycles).
We can make the structure of this billiard model more resembling to that
of Hill's problem by smoothing out the corner where the disks meet: then
there is a third fixed point which can be elliptic if the curvature of 
the dip is small enough so that the resulting stable island provides the
third structural element in the organization of the scattering orbits.

\acknowledgments

This work has been started as part of a research project financed 
by the Foundation for Fundamental Research on Matter in The Netherlands 
(project number: 93BR1099). 
Partial financial support from the Hungarian Scientific Research 
Foundation, under grant numbers OTKA F4286, F17166 and T17493 
is also acknowledged.

\pagebreak


\begin{references}
\bibitem{remark}
  These numbers differ slightly from those given in 
  Ref.~\cite{Petit-Henon} for the edges of the irregular domain
  due to the fact that we used a lower-order approximation in matching
  initial conditions to circular asymptotic solutions of Hill's equations.
  The difference (basically a shift in the order of $\eta_0^{-1}$)
  shows up in all $h$ values cited in the present paper; however,
  since we are interested in the structural properties of scattering 
  rather than actual numbers, this has no relevance for our findings
  presented here.
\end{references}
\end{document}